# Pseudo-Degeneracy in Handbell Modes

*Sungdo Cha*
Department of Physics, Kangwon National University, Chuncheon, Kangwon-do, Korea 200-701

*John Ross Buschert, Daniel King*
Department of Physics, Goshen College, 1700 South Main Street, Goshen, Indiana 46526, USA

## Abstract

The vibrational modes of handbells exist in nearly-degenerate pairs which have been split by nonuniformities in the bell. The possibility of intentionally re-combining a pair into pseudo-degeneracy was explored by adding mass to various locations on a C4 handbell. Depending on the location and mass added to the bell, the modes can be shifted in position and frequency as observed by a real-time holographic interferometry system. By moving and varying the added mass, one pair of modes was successfully brought into a pseudo-degeneracy.

## Ⅰ. INTRODUCTION

Like most bells, handbells have a rich set of vibration modes. Interestingly, each mode is split into a pair of nearly degenerate modes. The basic acoustics of handbells have been extensively studied[1,2] and the important modes of vibration are well characterized yet they remain an excellent source of interesting experiments for students[3]. In particular, less work has been done to characterize and manipulate the degenerate pairs of modes.

The aim of this experiment was to manipulate the individual members of one nearly degenerate pair of modes and to see whether it was feasible to bring the frequencies together to establish a pseudo-degeneracy. This would be attempted by adding a small point mass to the bell and varying the amount and the position of this mass. Adding mass to a handbell at a single point can change the frequencies of the various modes of vibration. However, due to the orientation of the modes on the bell this does not affect both members of a pair of modes equally. This offers the possibility of restoring to a particular mode a kind of degeneracy which has been broken by non-uniformities.

This paper will begin with a review the basic physics of handbell vibration modes. Then an experimental apparatus is described for measuring the frequencies and angular orientations of vibration mode pairs. This is followed by the measurements and results of the experiment attempting to restore the degeneracy of a particular pair of modes.

## II. Background

When handbells are rung, a great many vibration modes are excited. Figure 1 shows a handbell spectrum and exaggerated diagrams of a few of the vibration modes. The various modes are indexed according to the number of diameter-like nodal lines and the number of nodal circles. Typically only the (2,0) and (3,0) modes are tuned during the manufacture of the handbell to give it a sense of pitch. The other modes are not in harmonic relationships but contribute to the bell-like tone color. Relatively recently, handbells with many tuned harmonics have been made but they are not yet widely available.[4]

All the vibration modes in most types of bells come in pairs. The two modes of each pair have the same shape and nearly the same frequency but differ in orientation on the bell.[5] Fig. 2 shows the two (3,1) modes as imaged by holographic interferometry. The brightest white lines are nodal lines where the bell is moving very little. The bulls-eye patterns are anti-nodes where the amplitude is large. The naming system used in this research labels the lower frequency mode the ‘A’ mode and the higher frequency the ‘B’ mode. As is always the case, the two modes are orthogonal -- the anti-nodes of the (3,1)A mode line up with the nodes of the (3,1)B mode.

These two modes would be identical in frequency and truly degenerate if the bell were perfectly symmetrical about its axis. Real bells have nonuniformities, principally material variations and even voids from the casting process.[6] These imperfections cause a splitting of the symmetry of the degenerate modes into nearly degenerate pairs.

The split pairs typically differ by a few Herz which can lead to undesirable beating (sometimes called warbling) when both modes of a pair are excited. Bell makers try to position the clapper to strike at a nodal line of one mode to eliminate this beating. Striking on a nodal line does not excite a mode but will excite its twin which has an antinode there. Since there are so many pairs with different orientations, this cannot solve the problem completely and in fact, a significant fraction of finished bells are rejected due to undesirable beating. In very large carillon bells, these beating patterns can result in different sound patterns for listeners in different positions.[7]

It is possible by various means to drive the bell at a specific frequency. If the driving frequency is swept slowly through the two resonances of a pair, each one comes and goes but remains in a fixed orientation on the bell. Driving the bell at different locations changes the response amplitude of the modes but still does not vary the orientation of the nodal lines of each mode. The two modes of a pair stay in a fixed orientation on the bell which is determined by the particular nonuniformities of the bell.

Figure 3 is a graphical representation of the azimuthal angular positions of several mode pairs of a C4 handbell. Looking at the many different pairs of modes one finds no discernable pattern, they are not all aligned with each other in any clear way. Apparently the nonuniformities are not simple and they affect the various modes differently.

This figure also illustrates the problem of clapper position. Placing the clapper at 45 degrees would eliminate the (2,0)A mode (which has a nodal line there) but would still excite both (3,0) modes. So this study looked at a different approach to solving the beating problem by trying to restore the degeneracy in a particular pair of modes.

## III. EXPERIMENTAL SETUP

A Malmark C4 handbell was used for these investigations. The handle was removed and the bell mounted directly onto a rigid post attached to a machining turntable. Thus the bell could be readily rotated about its axis. The basic design of the experimental setup is shown in Fig. 4.

To make the bell scatter light better for the interferometry, we applied a fine layer of white spot-check developer.[8] Intended to help mechanics search for cracks in engines, this product is very easy to apply in a thin uniform layer and is removed by simply wiping it off with a cloth. While it does add slightly to the mass of the bell and it lowers the mode frequencies a few tenths of a Herz, it affects the two modes identically and so did not affect these results.

The modes were observed using a holographic interferometry system supplied by Stetson Associates.[9] In this real-time interferometry system, a reference beam and object beam converge and interfere in a digital camera. Multiple images with different phases of the reference beam are analyzed by an electronic system to recreate the holographic image.[10] The final image displayed on a monitor shows interference fringes due to any motion of the object.

When a handbell is rung in the usual way, the impact of the clapper excites dozens of modes of vibration simultaneously. It can instead be forced into a specific mode of vibration by driving it at a certain resonance frequency. A common method used in the past has been to attach a magnet to the bell and then drive it with a coil placed nearby. This has the disadvantage that the driving point cannot be easily moved. Worse for this study, the driving magnet would have added its own nonuniformity to the bell. Instead, the bell was driven by a unit which contained both a strong magnet and coil together. This little known method uses a permanent magnet to create a force on eddy currents generated by the coil. The resulting coupling between electric and magnetic fields in the bell and the driving coil effectively drives the bell at the frequency of the generator. Thus the unit can drive any conducting surface without contact or modification so nothing needed to be added to the bell. The signal source was an Agilent 33120A Function Generator amplified by a Crown D-75 Audio Amplifier operated in bridge mode to double the voltage. The frequencies and angular positions of the node lines for the first 15 pairs of modes were recorded. To accurately determine the angular orientation of the vibration modes was an important measurement for this research. A turntable was used to rotate the bell until a nodal line was centered on the hologram monitor image and then the angle of the turntable was noted.

The masses added to the bell were a series of tiny neodymium magnets placed on the inside and outside of the bell holding each other in place. All of the following studies focused on the (2,0) mode. This is the lowest mode of the handbell and is the same as the perceived pitch of the bell.

A coordinate system was established on the bell for recording the angular position of the driver and of the nodal lines of the two modes. The system was chosen so that the A-mode (the lower one of the pair) had an antinode at the zero of the scale. We referred to this as the azimuthal angle on the bell. Later, when the vibration modes shifted, however, this azimuthal scale was kept fixed to the bell.

## IV. MEASUREMENTS AND RESULTS

In the first experiment, a mass of 2.96 g was placed at various azimuthal locations around the bell while keeping the vertical position the same at 4 cm from the rim of the handbell. Fig. 5 shows the result on the frequencies of the two modes. The frequencies of the two modes without the added mass were 261.54 Hz and 262.46 Hz. When the mass was added at the antinode of the A mode (0 degrees) its frequency dropped by 1.02 Hz while the B mode dropped by only 0.14 Hz. Then as the mass was moved around the bell the B mode dropped while the A mode rose. At the angle of 45 degrees, the A mode had nearly returned to its original frequency and the B mode had dropped by 0.93 Hz. This angle is where the B mode has its antinode and the A mode has a nodal line. This shows the expected result that the added mass has the greatest effect on a particular mode when it is located in the center of the antinode, the region of greatest motion.

A more interesting effect occurs in the azimuthal positions of these modes as the mass is moved around (fig. 6). The important result here is that when the added mass is placed on a nodal line of either mode, there is no rotation of the modes, only the frequencies are changed. When the added mass is not on a nodal line, both modes rotate together and remain orthogonal. They rotate in a direction that would bring the antinode of the lower mode nearer to the added mass. It is as if the added mass drags the lower mode around with it to some degree.

Having determined the effects of putting mass in different locations, different masses were added to a fixed location. The location chosen was at the antinode of the B mode in order to primarily drop the higher of the two modes. Thus by varying the amount of mass the two frequencies could be brought together. As seen in Fig. 7, the two modes converge in frequency as added mass is increased. They cross at about 3.7 g and begin to diverge again. One way to describe this would be to say that the nonuniformities in the bell which cause the splitting are being perturbed by the added mass up to the crossing point. After that point, the added mass has the larger effect and it could be said to cause the splitting while the non-uniformities in the bell perturb it with the lesser effect.

Through most of this experiment, the A and B modes were easily identified by the orientation of the mode which remained constant. However, at the exact point of crossing it was not so clear. When the frequencies were brought within about 0.1 Hz, it was no longer possible to excite just one or the other by adjusting the frequency because the resonances overlapped. So both modes were being excited and the angular position of the resulting pattern was in between the positions of the two pure modes. The graph suggests that if precisely the right mass and position is chosen, the two modes should have exactly the same frequency and be virtually indistinguishable. In this case, when a mass of 3.71 g was placed at the B mode nodal line and 3 cm from the rim of the bell, the two modes were be brought to the same frequency, 261.19 Hz.

The effect of varying amounts of mass was found to be linear. Thus it would be possible to find the right mass for the mixed mode by measuring only the frequency difference without the added mass and the frequency difference with too much added mass. The right mass can be found assuming a linear dependence and finding the zero crossing value.

One question that comes up is what has happened to the many other split modes of the bell. It might be hoped that these also would be brought back together. But this is not the case in general because the nonuniformities are complicated and the various modes are not aligned. This mass placement was on the antinode of the (2,0)B mode in order to lower it and combine it with the (2,0)A. But this location is far from the antinode of the (3,0)B (refer to figure 3) and would actually widen the splitting of that pair.

Figure 8 shows two plots of vibration amplitude versus driving frequency. One plot was made with the added mass to create the mixed mode and the other was without any added mass. With no mass the two modes are well separated but with the suitably chosen mass they have coalesced into one peak which has the same width (FWHM, 0.09 Hz).

To determine whether these coalesced modes really behave as a degenerate pair, the bell was rotated by increments while keeping the driver fixed to the table. When the modes are separate (no added mass), only one mode can be excited at a time. They remain fixed on the bell and rotate with it regardless of the location of the driver. With the added mass bringing them together, the driver is simultaneously exciting both modes. As the bell rotates, the driver's position varies with respect to each mode and so the amplitude of each one varies. But the resulting oscillation is as if there were only a single mode rotating on the bell but staying fixed with respect to the driver. This is just what would happen if the modes were truly degenerate. The results of rotating the bell with and without the added mass can be seen in figure 9. A dark line just visible on the top of the bell shows its orientation in each picture. The bell is rotated 20 degrees between successive images.

In the series without the added mass, the mode pattern clearly rotates with the bell. The second case has just the right added mass to bring the two modes to the same frequency. The pictures give the appearance of an oscillation mode that stays fixed relative to the driver. This pseudo-degeneracy has not eliminated the asymmetry of the bell, it still has nonuniformities. These not only split the mode's frequencies, they could also change the mode shapes slightly away from the apparent symmetry they seem to possess. Adding mass creates another non-uniformity to bring the frequencies back together. While the two modes may still be distinct and still slightly asymmetric, they are acting in nearly all measurable respects like a degenerate pair.

Figure 10 shows there is another way of seeing the resulting pseudo-degeneracy. Here the bell is driven at a resonant frequency and the amplitude of the vibration mode is graphed as the bell is rotated. With no added mass, the modes are well separated in frequency and the bell can be driven at the frequency of only one mode at a time. When this is done, the A- or B- mode comes and goes according to the whether the driver is exciting it at a node or an antinode. In the pseudo-degenerate case however, there is only one frequency. When the bell is driven at that frequency the amplitude is nearly constant as the bell rotates. .

## V. CONCLUSIONS

An experimental system for the investigation of frequency and azimuthal position of handbell vibration modes has been demonstrated. The system is particularly suited to the study of twin modes. This system was used to study the effects of placement of point masses on the frequency and azimuthal position of these twin modes which have been split by nonuniformities in the bell. By suitably choosing the placement of the added mass it has been shown to be possible to manipulate the individual members of a twin pair. Using this ability a pseudo-degeneracy of one pair of twin modes has been successfully restored as shown by holographic interferometry photos. The method is only suitable to solve one splitting and the effort needed for this achievement is not likely to be cost-effective in the mass production of most handbells. Combined with other methods however, it might be a useful tool in the fine tuning of larger handbells or of carillon bells.

## ACKNOWLEGEMENTS

This study is partially supported by Kangwon National University.

## References

1. Thomas D. Rossing and Robert Perrin, “Vibrations of bells,” Appl. Acoust. **20**, 41-70 (1987).

2. T. D. Rossing “Acoustics of Bells” American Scientist, 72, 440-447 (1984)

3. Kristen Menou, Benjamin Audit, Xavier Boutillon, and Holger Vach, “Holographic study of a vibrating bell: An undergraduate laboratory experiment” Am. J. Phys. **66** 380 (1998)

4. N McLachlan, BK Nigjeh, A Hasell, “The design of bells with harmonic overtones,” J. Acoust. Soc. Am. Volume 114, Issue 1, pp. 505-511 (July 2003)

5 Ralph T. Muehleisen and Anthony A. Atchley, “Fundamental azimuthal modes of a constricted annular resonator: Theory and measurement,” J. Acoust. Soc. Am. **109**, 480–487 (2001).

6 2 R. Perrin, G. M. Swallowe, T. Charnley and C. Marshall, “On the debossing, Annealing and Mounting of bells,” J. Sound Vib. **227**, 409-425 (1999).

7 3 Seock-Hyun Kim, Chi-Wook Lee and Jang-Moo Lee, “Beat characteristics and beat maps of the King Seong-deok Divine Bell, ” J. Sound Vib. **281**, 21-44 (2005).

8 Mechanics use several sprays to find tiny cracks in engine blocks. The last component is a very fine white powder that is also ideal for making shiny objects show up better for holography. The product we used, Magnaflux Spotcheck Developer SKD-S2 is available online from markingpendepot.com.

9 Karl Stetson Associates, LLC 2060 South Street Coventry, CT 06238 USA

10 Karl A. Stetson and William R. Brohinsky “Electrooptic holography and its application to hologram interferometry,” Applied Optics, Vol. 24, Issue 21, pp. 3631-3637 (1985)

## Figures and Captions

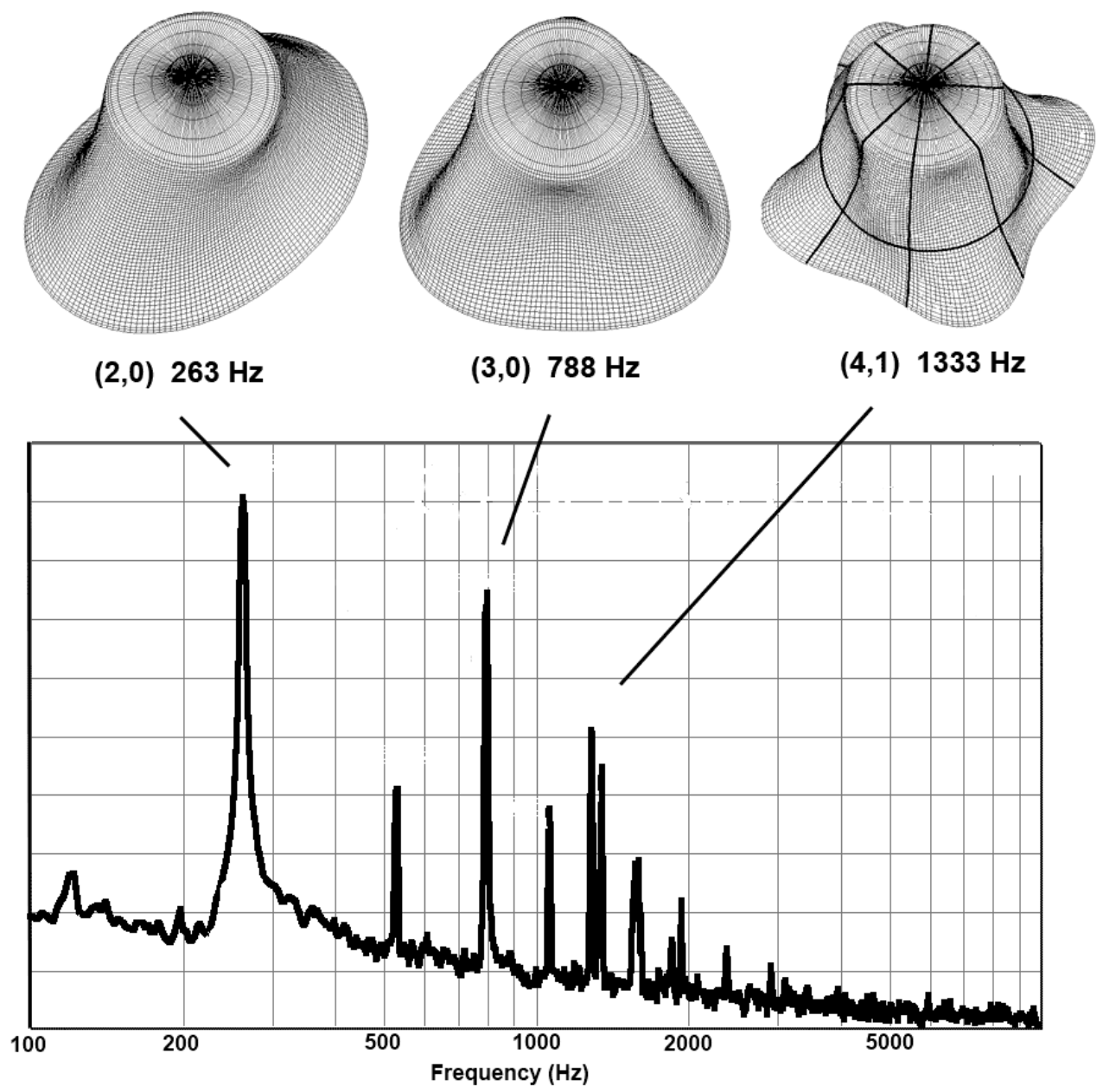

FIG. 1 The sound spectrum of a C4 Handbell. Above the graph are exaggerated views of the vibration mode responsible for three of the peaks. The (4,1) mode has the four nodal diameter lines and the one nodal circle drawn on it to illustrate the indexing scheme of the modes.

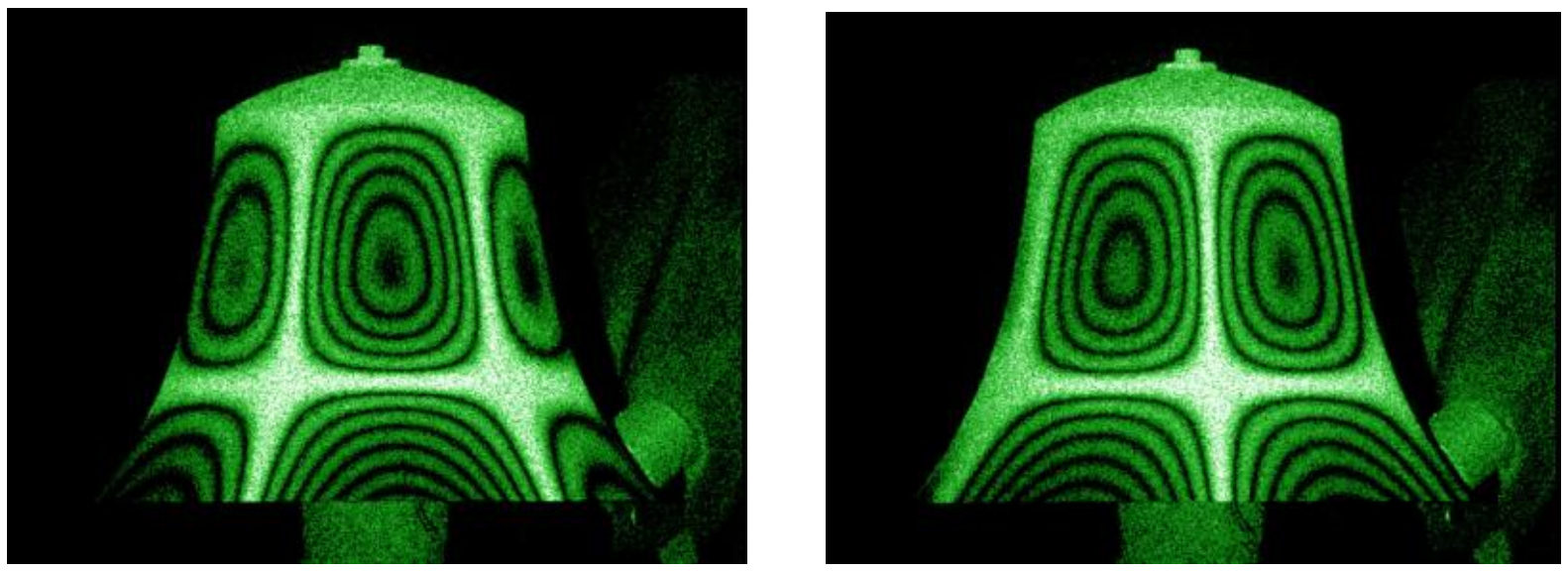

FIG. 2. The two (3,1) modes in C4 handbell imaged by holographic interferometry. The left image is the (3,1)A at 1239.8 Hz and the right image is the (3,1)B at 1240.5 Hz.

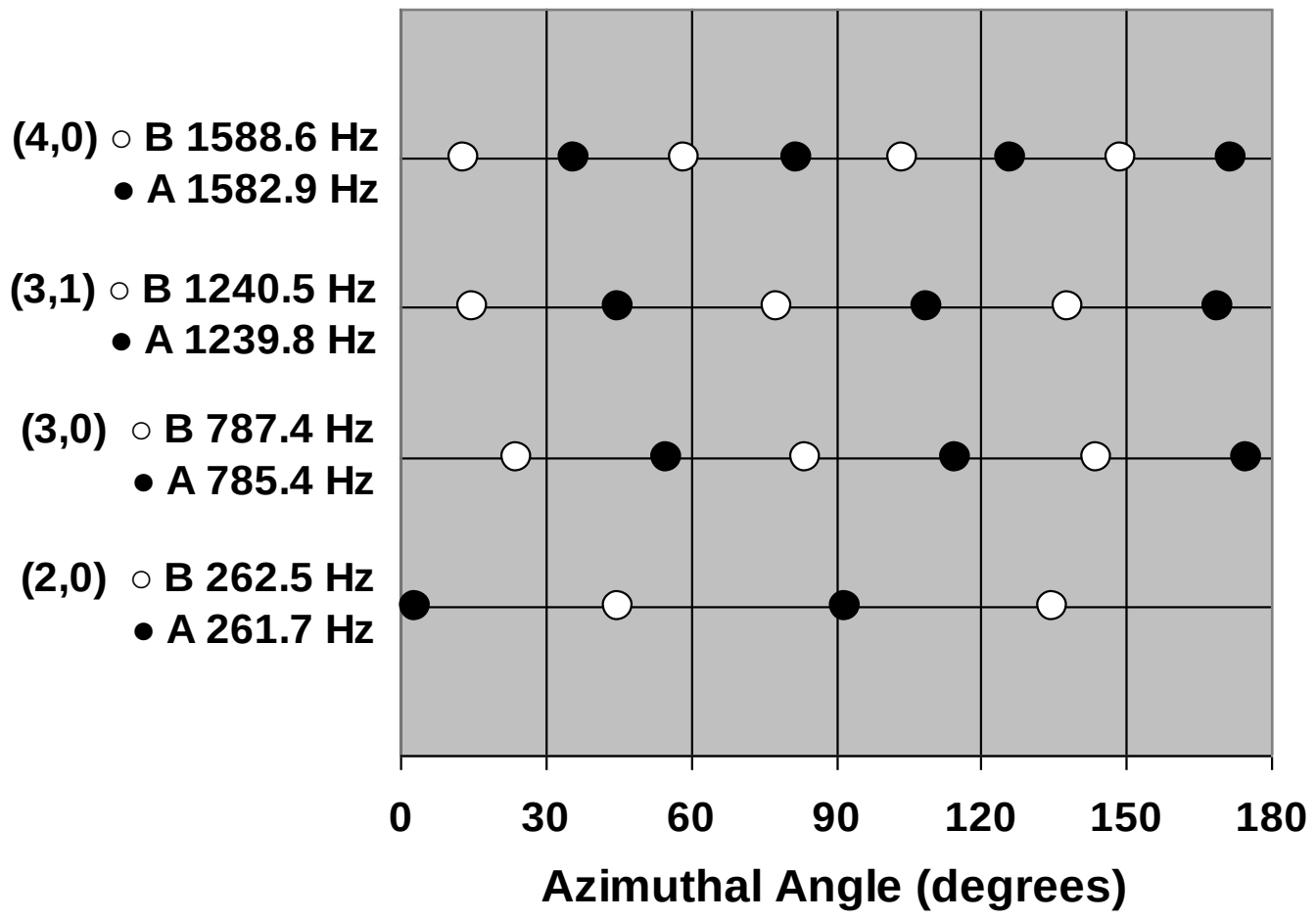

FIG. 3. Orientation of pairs of vibrational modes on a C4 handbell. Each modal pair is plotted on a horizontal line. The symbols mark the azimuthal positions of the antinodes with repect to a fixed mark on the bell. The A and B modes are always orthogonal but differently indexed modes are not aligned with each other.

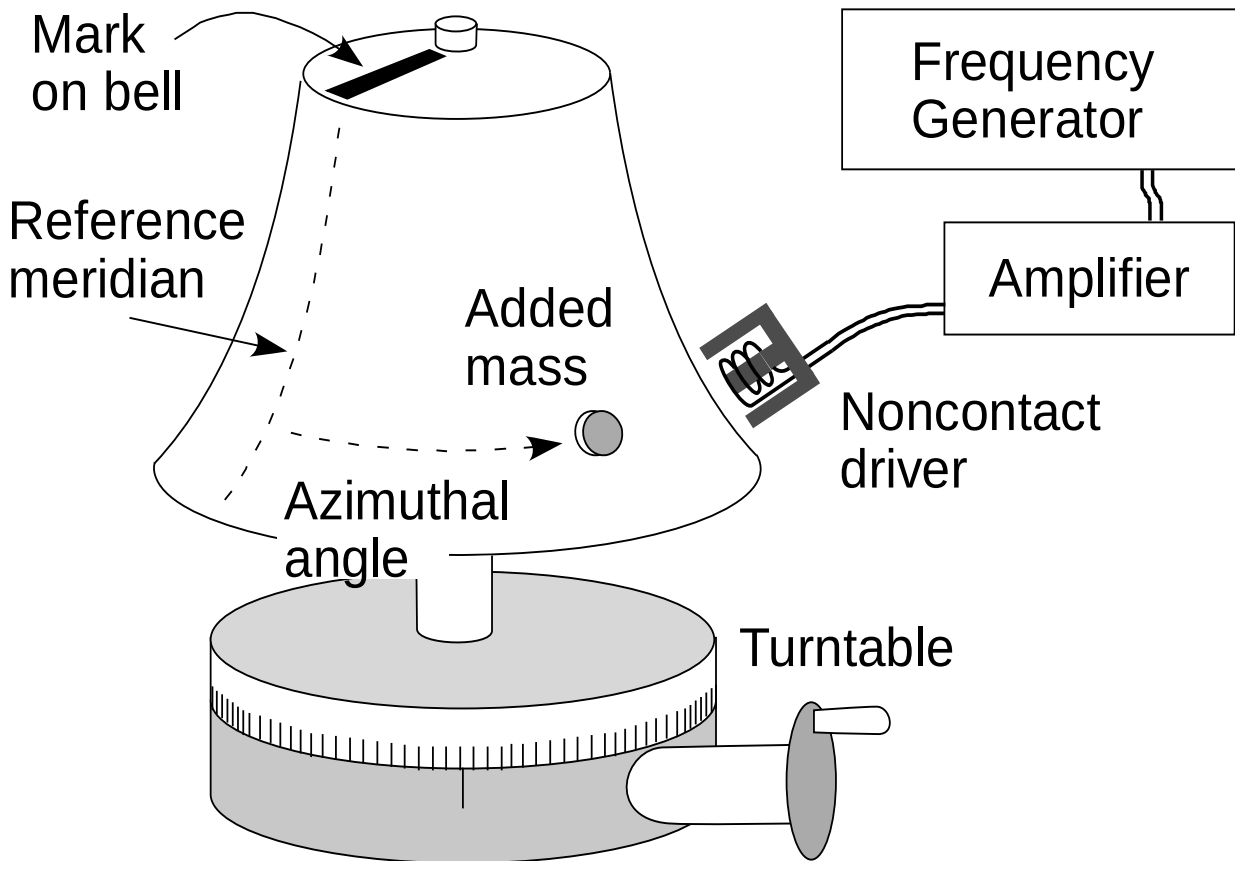

FIG. 4. Experimental Setup. A signal generator drives the bell at a particular frequency through a noncontact driver. The bell could be rotated on the turntable while being imaged using a holographic interferometry system. Vibration mode features and added masses were located on the bell by the azimuthal angle from a reference meridian which was fixed on the bell.

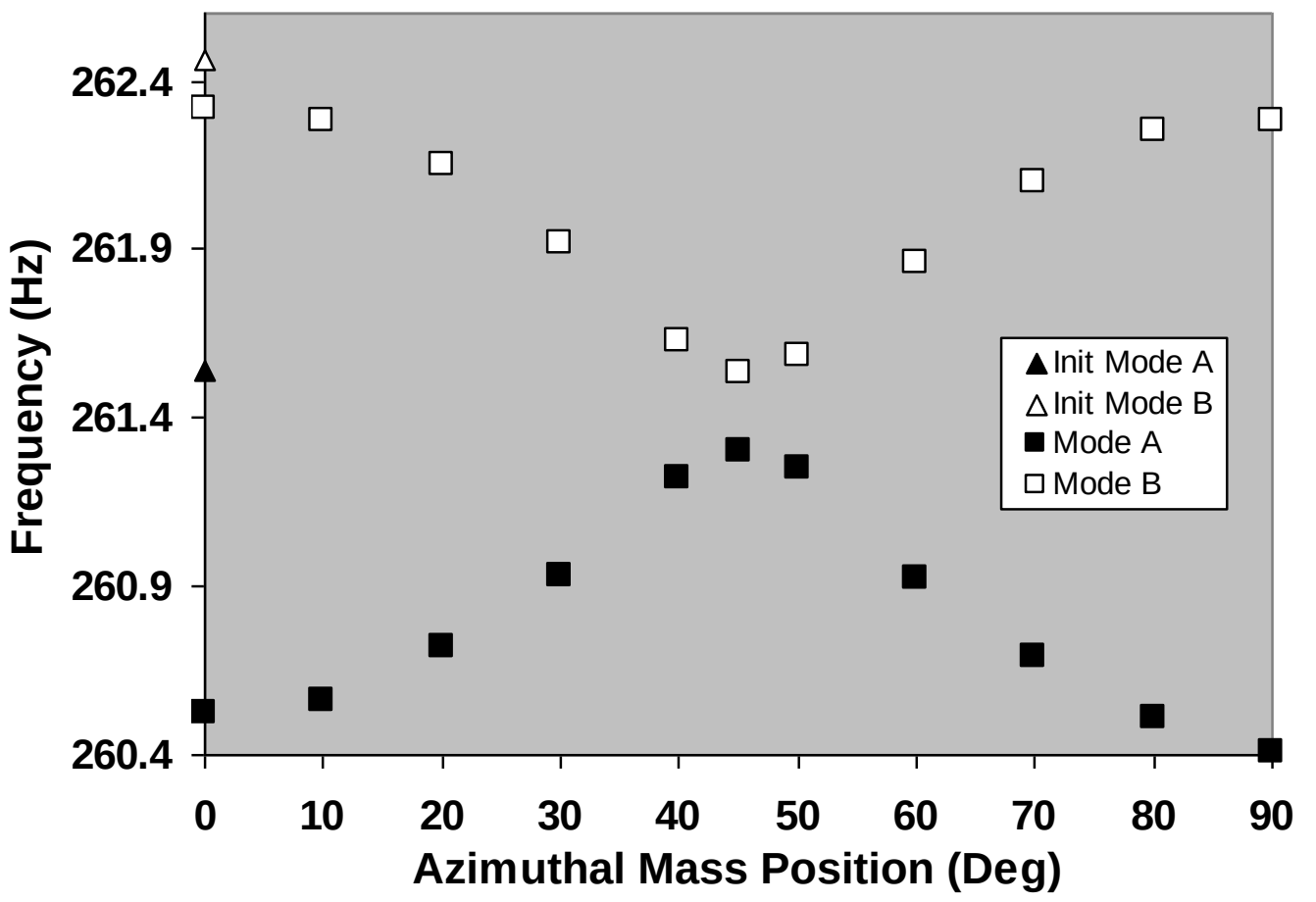

FIG. 5. Frequency change versus azimuthal position of the mass. On the left axis the triangular symbols represent the original frequencies of the two modes without added mass.

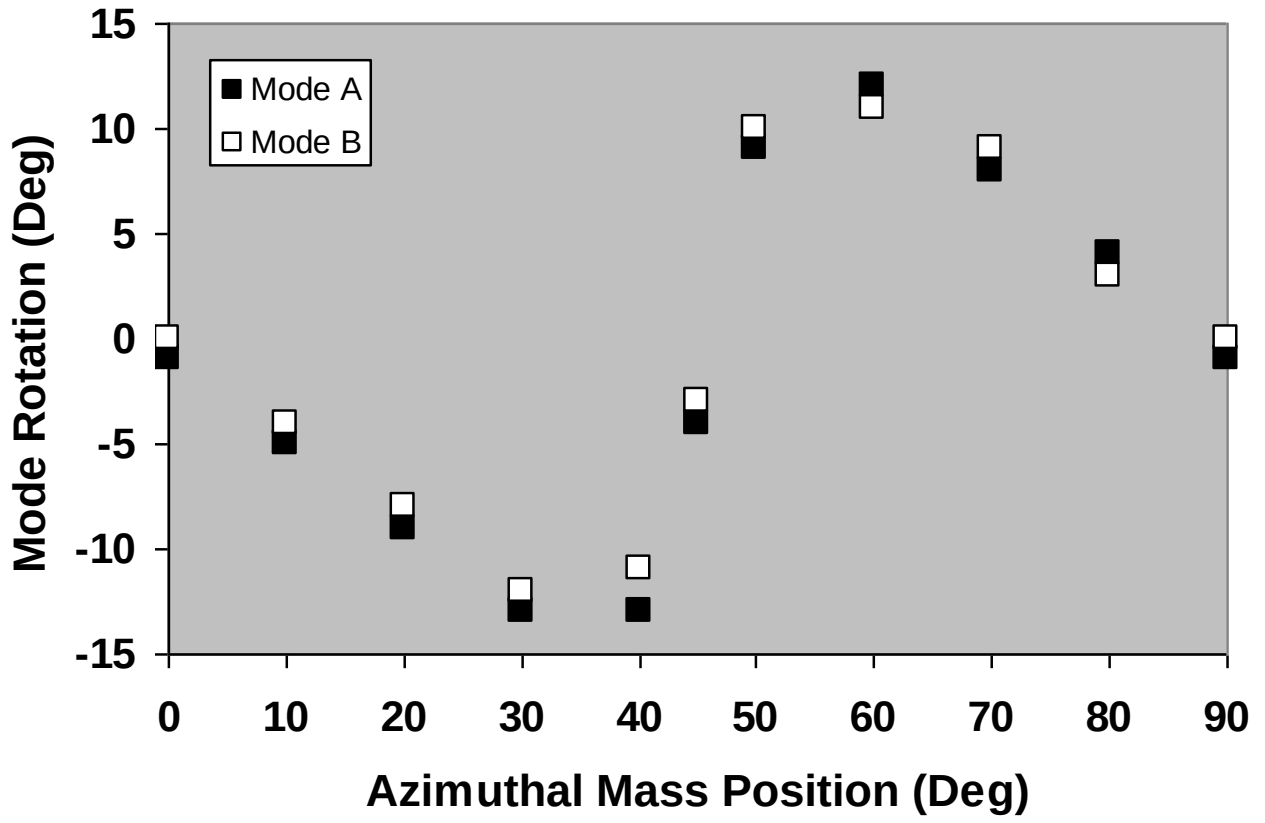

FIG. 6. Angular shift of the modal pattern as a function of the location of the added mass.

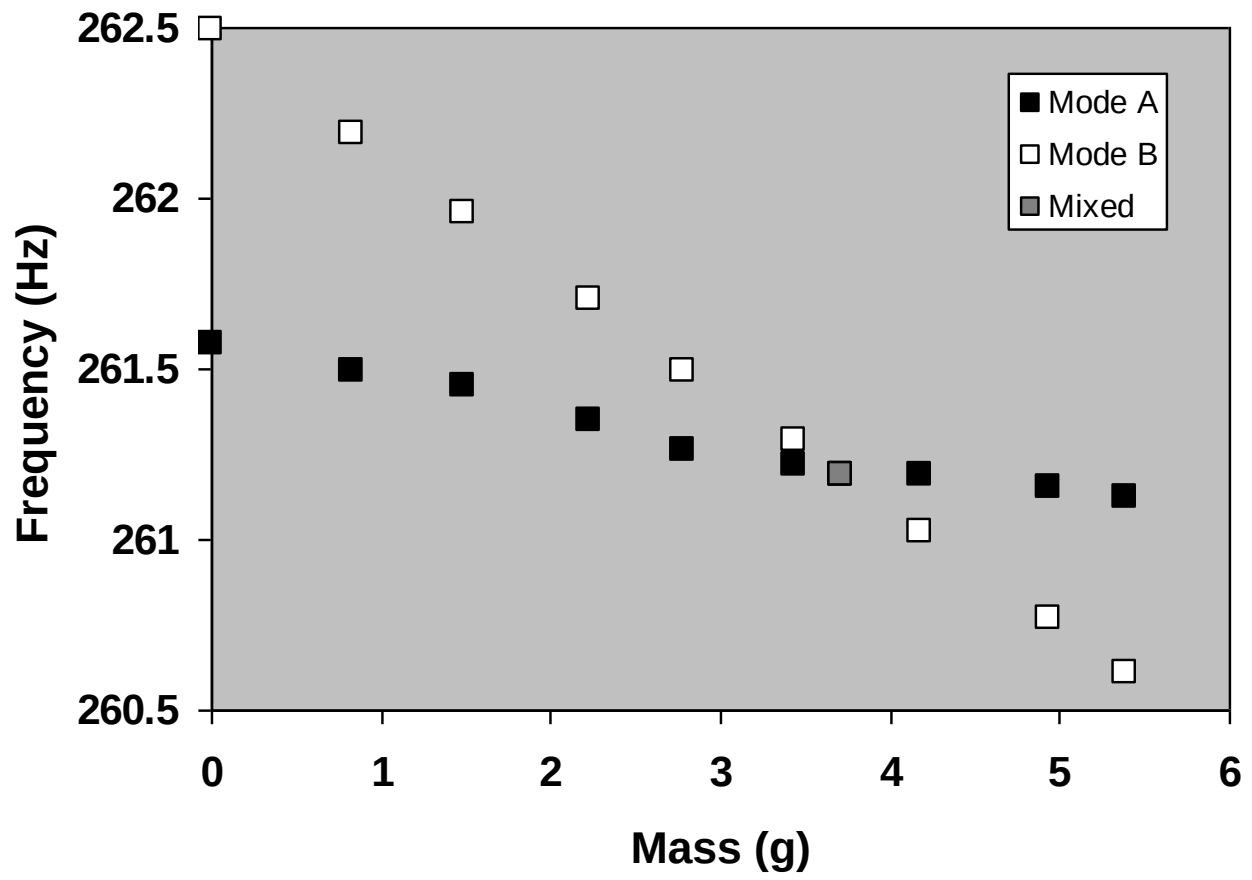

FIG. 7. The effect of different amounts of added mass on the frequencies of the A- and B-modes. At 3.7 g the two modes become difficult to distinguish.

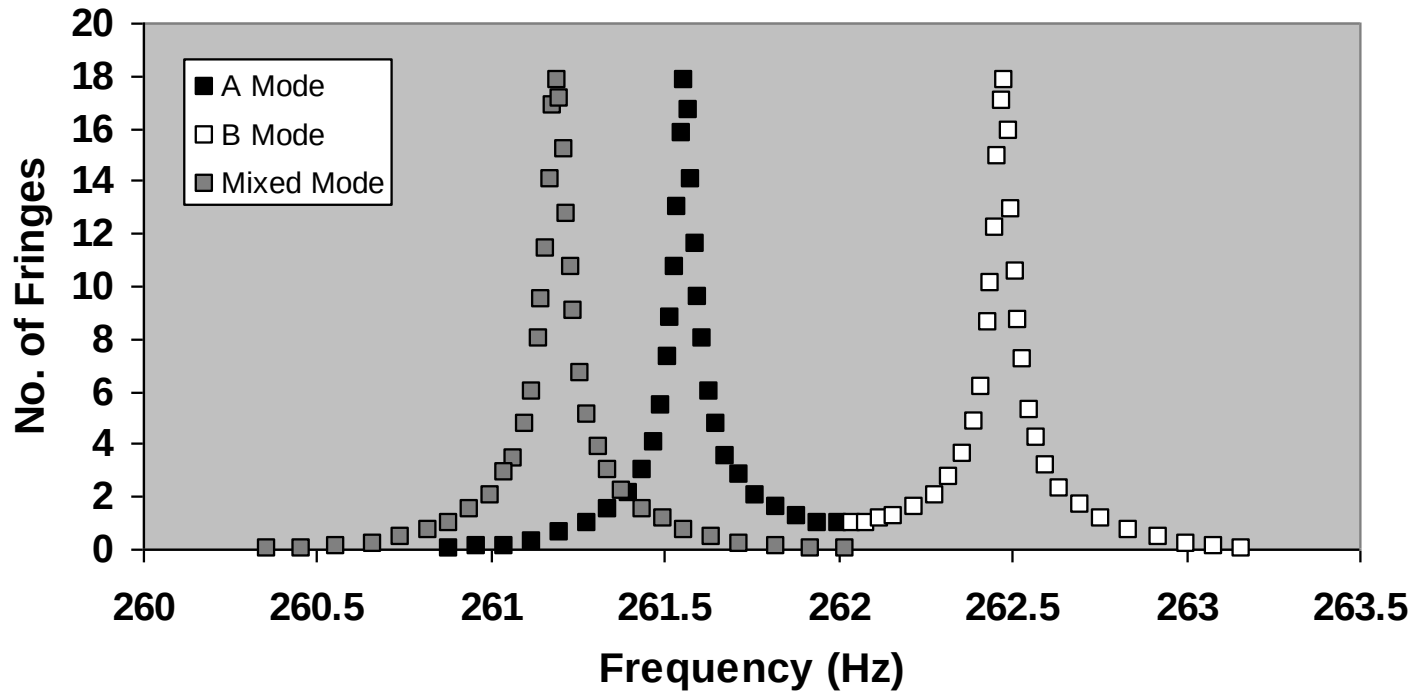

FIG. 8. The amplitude of the vibration modes versus the driving frequency. Black and white symbols are the two modes without added mass. The gray symbols are for the combined mode with added mass.

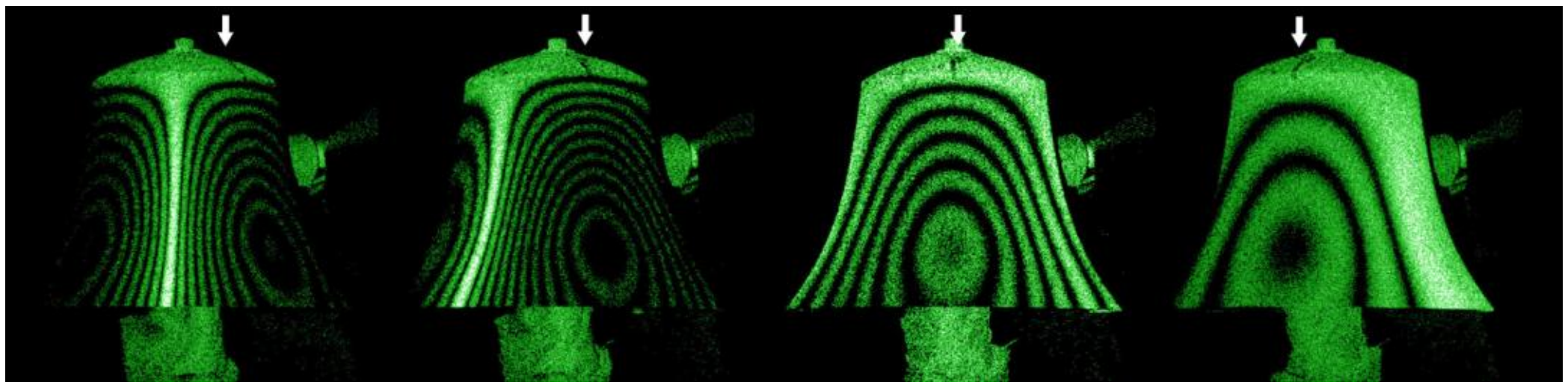

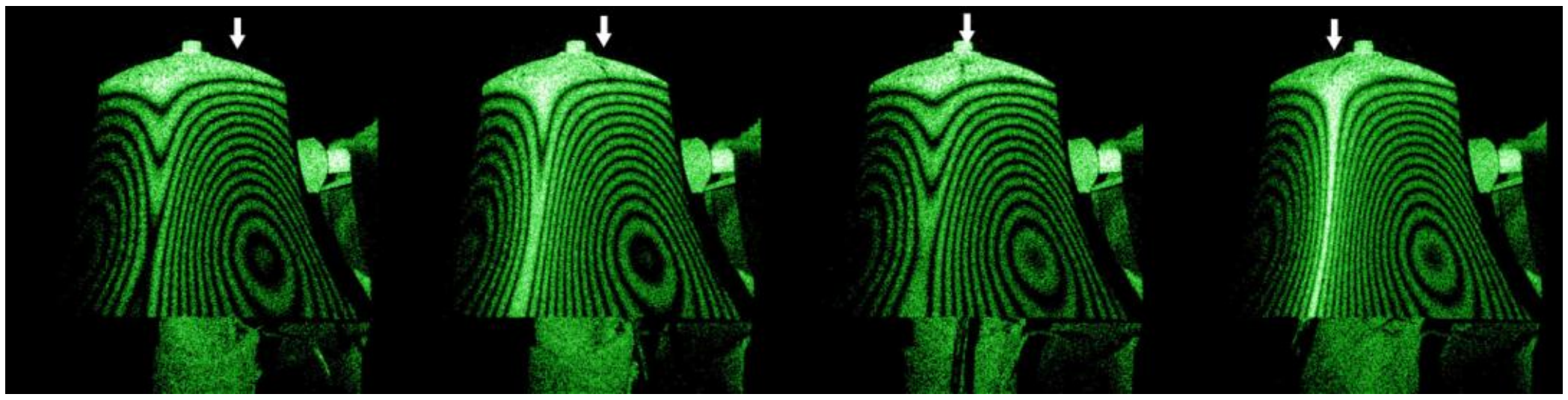

FIG. 9. Two series of holographic images of (2,0) mode. In each successive image, the bell is rotated to the left by an increment of 20 degrees . In the upper series, there is no added mass and so the vibration pattern moves with the bell. In the lower series, a mass has been added which brings the split modes together. In this case the vibration pattern stays in essentially the same position from one image to the next.

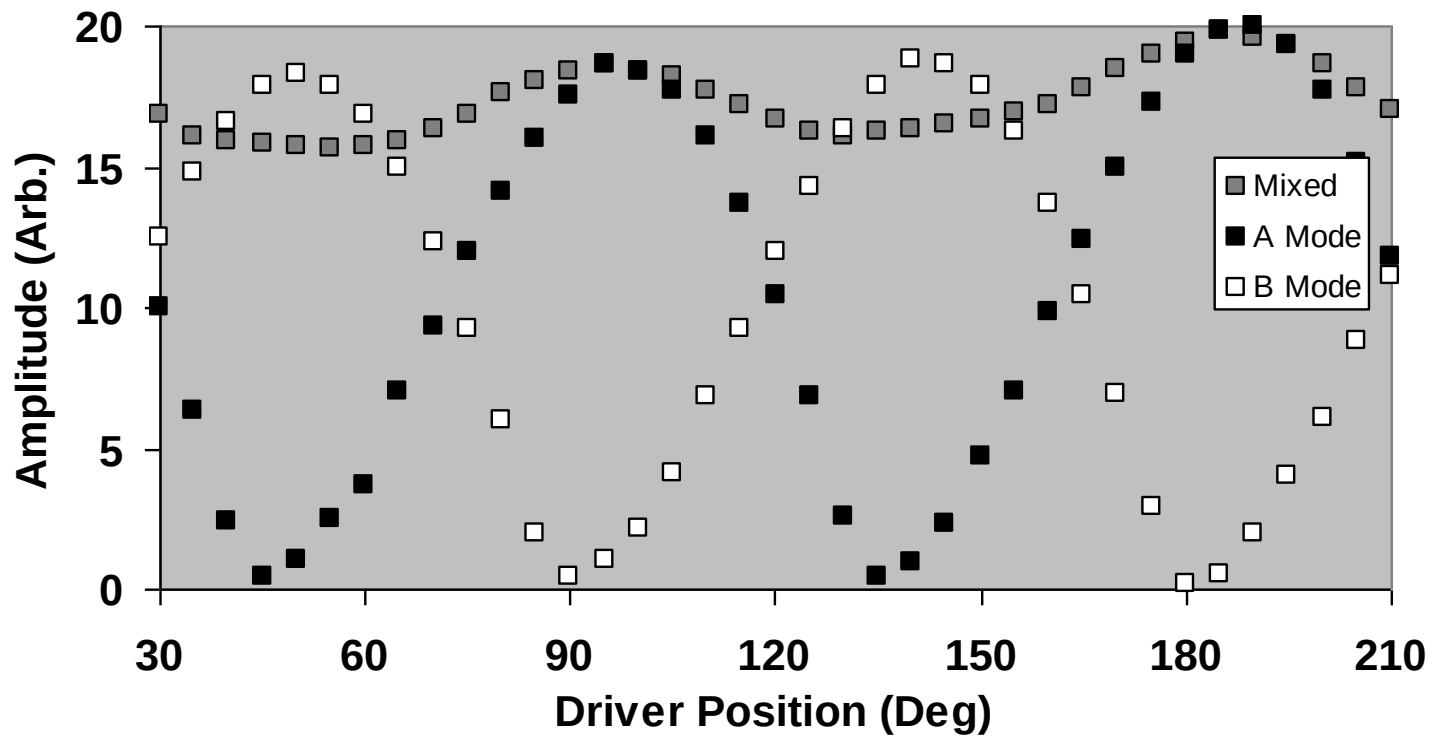

FIG. 10. The amplitude versus driver position of the A- and B-modes without added mass and of the combined mode.